# Endo-microscopy beyond the Abbe and Nyquist limits

Lyubov V. Amitonova[1,2,*] and Johannes F. de Boer[1]

[1]*LaserLaB, Department of Physics and Astronomy, Vrije Universiteit Amsterdam, De Boelelaan 1081, 1081 HV Amsterdam, The Netherlands*
[2]*Advanced Research Center for Nanolithography (ARCNL), Science Park 106, 1098 XG, Amsterdam, The Netherlands*
[*]e-mail: l.amitonova@vu.nl,

**Abstract**

For several centuries, far-field optical microscopy has remained a key instrument in many scientific disciplines including physical, chemical and biomedical research. Nonetheless, far-field imaging has many limitations: the spatial resolution is controlled by the diffraction of light and the imaging speed follows Nyquist-Shannon sampling theorem. The recent development of super-resolution techniques has pushed the limits of spatial resolution. However, these methods typically require complicated setups, long acquisition time and are still not applicable for deep-tissue bioimaging. Here we report imaging through an ultra-thin fiber probe with a spatial resolution beyond the Abbe limit and a temporal resolution beyond the Nyquist limit simultaneously in a simple and compact setup. We use the random nature of mode coupling in a multimode fiber, the sparsity constraint and compressive sensing reconstruction. The new approach of super-resolution endo-microscopy does not use any specific properties of the fluorescent label such as depletion or stochastic activation of the molecular fluorescent state and therefore could be used for label-free imaging. We demonstrate a spatial resolution more than 2 times better than the diffraction limit and an imaging speed 20 times faster than the Nyquist limit. The proposed approach could significantly expand the realm of the application of nanoscopy for bioimaging.

**Introduction**

Recent years have witnessed the development of super-resolution far-field fluorescence microscopy that allows the diffraction-limited resolution to be surpassed, unveiling processes at the nano-scale level[1]. Structured illumination microscopy (SIM) utilizes spatial modulation of the fluorescence emission with patterned illumination[2]. However, it yields an improved resolution only by a factor of two. Stochastic optical reconstruction microscopy (STORM) and photoactivation localization microscopy (PALM) are based on stochastically switching individual molecules on at different times[3,4]. Stimulated emission depletion microscopy (STED) increase resolution through shrinking the point-spread function (PSF) by depleting the fluorescence emission in the periphery of the diffraction-limited spot[5]. These techniques cannot be used with any dye due to specific requirements to the fluorescent labels.

Achieving super-resolution *in vivo* deep in living tissue is extremely challenging due to limited optical access, optical aberrations, and low imaging speed. The maximum demonstrated imaging depth for super-resolution remains only 120 μm below the tissue surface[6]. For deep-tissue microscopy, minimally invasive endoscopes are widely used. The recent emergence of the spatial wavefront shaping approach[7–9] allowed a conventional step-index multimode (MM) fiber to be utilized as an ultra-thin aberration-free imaging probe[10–14]. The numerical aperture (NA) of MM fibers approaches 0.9[15,16], enabling high-resolution imaging. Recently, we demonstrated super-fast imaging through a MM fiber[17]. Simultaneously, methods to enhance the resolution of multimode fiber-based imaging are rapidly developing[18,19]. However, it is still a big challenge to incorporate super-resolution in MM fiber approaches.

Parallel efforts aim to improve the imaging speed[20]. Not surprisingly, advanced methods of diffraction-unlimited resolution are time-consuming and the trade-off between spatial and temporal resolution affects all super-resolution techniques. Most approaches require point-by-point scanning and, consequently, the acquisition speed is limited by the Nyquist-Shannon sampling theorem[21].

Here we propose and experimentally demonstrate fluorescence imaging through a thin MM fiber probe with a spatial resolution beyond the Abbe limit and a temporal resolution beyond the Nyquist limit, meaning super-resolution and super-speed at the same time. The proposed approach doesn't put any specific requirements on the fluorescent labels except brightness and could be used with any standard dye with reasonable concentration. We exploit the random nature of mode coupling in a MM fiber, the sparsity constraint inherent to any natural image and compressive sensing integrated into a compact fiber probe. The new approach potentially allows minimally invasive deep-tissue imaging with a spatial resolution more than two times better than the diffraction limit and an imaging speed 20 times faster than is required by the Nyquist-Shannon limit, opening a new avenue for *in vivo* ultra-fast nanoscale imaging.

## Results

**Super-resolution image reconstruction via compressive sensing**

The Abbe diffraction limit[22] is often considered as the theoretical limit of spatial resolution achievable by a conventional optical system. Nevertheless, it was shown that it is possible to computationally resolve infinitesimally small details[23]. Unfortunately, existing methods for bandwidth extrapolation[24–26] are known to be extremely sensitive to noise. As a result, the diffraction limit remains a practical resolution boundary of a simple imaging system[23,27].

Recently, compressive sensing – a novel imaging paradigm robust to noise in the measured data – has emerged[28,29]. Compressive sensing enables signal acquisition with a large reduction in sampling for signals that have a sparse representation, vastly reducing the number of measurements that need to be done[30–33]. It exploits the fact that many natural signals are sparse in some basis. Compressive sensing can be used to significantly reduce the number of measurements in endo-microscopy applications[17,34]. Remarkably, compressive sensing serves as an effective method for rejecting certain common types of noise, which typically do not have sparse transforms. As a result, super-resolution imaging by using sparsity constraint is theoretically feasible[35,36]. The idea of super-resolution imaging via compressive sensing was originally demonstrated by using artificial low pass-filtering to measured data[37,38]. Recently, the possibility of reconstructing images in the case when the illumination transverse size is bigger than the feature size of the sample was shown, however, with a resolution of only tens of micrometers[39,40].

The spatial resolution of the compressive sensing approach is fundamentally limited by sparsity. In an ideal noise-free oversampling scenario, the resolution improvement could be up to a factor of $1/(2\beta)$, where $\beta$ is the sparsity parameter (the relative fraction of nonzero basis functions occupied by the original image)[37,38]. However, it is important to note that the information doesn't have to be sparse in real space: any mathematical basis that is sufficiently uncorrelated with the sampling basis will work. The majority of natural objects are sparse in some basis. In the following sections we also consider the practical issues limiting the spatial resolution of the proposed super-resolution endo-microscopy such as noise, stability and orthogonality of the sensing basis, and the number of measurements.

**Principles of the optical setup**

The idea of MM fiber based ultra-fast super-resolution endo-microscopy is sketched in Fig. 1a. The system consists of three main components: a continuous-wave (cw) laser source with a scanning system, a MM fiber probe, and a single-point detector. Pump light is scanned across the fiber input facet creating different illumination patterns on the fiber output. The total fluorescent response from the sample is collected by the same fiber probe, propagated back and measured by the single point detector. The Methods section and Supplementary Fig. S1 present a complete description of the experimental geometry.

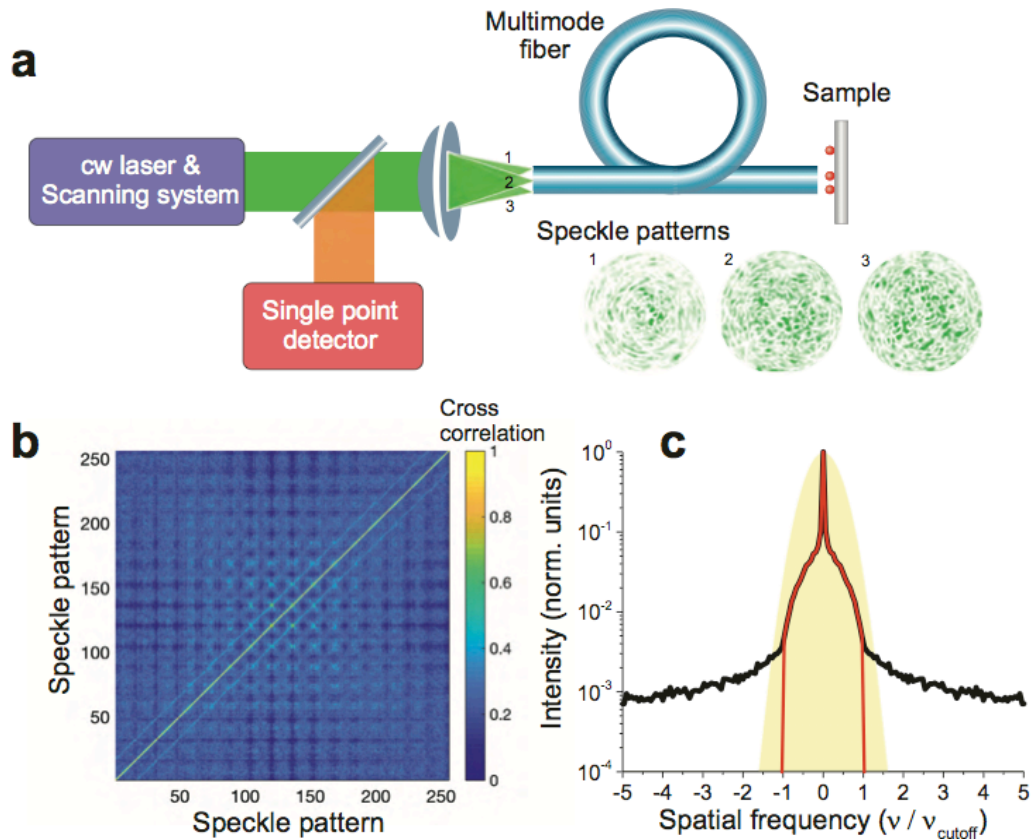

**Fig. 1 | The concept of the multimode fiber based super-resolution and super-speed endo-microscopy. a,** Experimental setup that consists of 3 main components: a continuous-wave laser source with a scanning system, a MM fiber probe, and a single-point detector. Pump light is scanned across the fiber input facet creating different illumination patterns on the fiber output. The total fluorescent response from the sample is collected by the same fiber probe, propagated back and measured by the single point detector. **b,** Cross-correlation coefficients of different speckle patterns generated in the MM fiber with NA = 0.22 for a total of 256 patterns. **c,** The one-dimensional Fourier spectra ($k_x$ components). Black and red lines represent the average spatial frequency spectra of speckle patterns generated in the MM fiber before and after low-pass filtering, respectively. The filled yellow area shows the spatial power spectrum of an ideal Gaussian beam. The cutoff frequency of the MM fiber was calculated as $v_{cutoff} = 2NA/\lambda$.

The key component of the proposed approach is a step-index MM fiber. We used two different fibers to demonstrate the independence of the proposed approach to the probe geometry. The first fiber has a silica core of 50 μm in diameter and an NA = 0.22. The Abbe diffraction limit is $\lambda/(2NA)$ = 1.21 μm. The second fiber has a core of 105 μm in diameter, an NA = 0.1, and an Abbe diffraction limit of 2.66 μm. Each MM fiber supports approximately 2000 modes[41] at a wavelength of 532 nm leading to the Nyquist limit of about 4000 measurements for diffraction-limited sampling.

**Sensing basis provided by mode scrambling in the multimode fiber**

One of the fundamental premises underlying compressive imaging is incoherence. It means that the sampling method should ensure the lowest correlation between any element of the sensing basis and any element of the representation basis. Consequently, to have freedom of choice of a representation basis we need a universal sensing basis. Random matrices are known to be largely uncorrelated with any fixed basis that makes them a very good tool for compressive sensing[28]. We create a universal basis for compressive endo-microscopy by using the random nature of mode coupling in the MM fiber[42]. The light field in a MM fiber (step-index circular waveguide) can be expressed as a finite sum of so called HE, EH, TE and TM modes propagating with their own speed. Different focal positions on the proximal fiber facet result in different speckle-like mode interference on the distal facet. An example is presented in Fig. 1a, where 3 locations of the input beam correspond to 3 different speckle patterns on the output. A total of 256 patterns were created by scanning a focused input beam over 256 points organized in a square lattice (16x16) on the input fiber facet. It is known that the full transmission matrix of a fiber can be measured using a similar scanning procedure[43], meaning that this experimental geometry potentially addresses all fiber modes.

Firstly, we characterized the orthogonality of the experimentally generated basis by calculating the correlation coefficient between any two measured speckle patterns (see Methods). The results for the first fiber are presented in Fig. 1b. The correlation between two random patterns generated in the MM fiber is close to zero, confirming the high level of orthogonality of our speckle basis. However, focal positions on the fiber facet located very close to each other lead to non-negligible correlations between the corresponding speckle patterns. As a result, the experimentally created basis is not ideal. We want to emphasize that in all our simulations, we use this experimentally measured non-ideal speckle-based point spread functions (PSFs) and demonstrate that the basis is sufficient for super-resolution endo-microscopy.

Secondly, we analyzed the spatial frequency content of the speckle patterns produced by the MM fiber with a cutoff frequency $\nu_{cutoff} = 2NA/\lambda$. The spatial power spectra was calculated by a 2D fast Fourier transform (FFT). The average one-dimensional spectrum ($k_x$ component) is presented in Fig. 1c by the black line. Experimental results are consistent with the theoretical predictions: the measured data at the frequencies beyond $\nu_{cutoff}$ are close to zero and effectively contain only noise. A zoomed-in camera image of the speckle pattern generated by the first MM fiber is presented in Supplementary Fig. S2a. Red arrows indicate the presence of dead pixels on the camera sensor that lead to high-frequency noise. To make sure that we do not introduce any high-frequency components in the simulations of our imaging system, we cut the Fourier spectrum of each measured speckle pattern at $\nu_{cutoff}$ (the red line in Fig. 1c). Low-pass filtered speckle patterns were recovered by the inverse 2D FFT. A zoomed-in example of the speckle pattern after low-pass filtering is presented in Supplementary Fig. S2b. There are no more high-frequency components preserved in the speckle PSF.

Finally, we measured the reproducibility and stability of the speckle patterns while the focused laser beam is switched between different locations on the fiber input. We scanned the focused beam over 100 different locations and recorded 100 different speckle patterns every 3 minutes. Then we characterized the reproducibility of the speckle patterns as the cross-correlation coefficient between the pattern measured at zero time and the pattern measured after the time delay and switching between different focal positions. We found out that during 45 minutes the averaged crossed-correlation coefficient is more than 0.95 confirming a high level of the reproducibility. The results are presented in Supplementary Fig. S4.

**Simulation with low-pass filtered speckle patterns: beyond the Abbe limit**

As we will now demonstrate, exploiting the disorder in MM fibers in combination with the sparsity constraint allows to create a compact ultra-fast super-resolution endo-microscope. As a sample, we use a structure consisting of parallel stripes in horizontal and vertical directions separated by 390 nm (Fig. 2a) and by 480 nm (Fig. 2b). Feature sizes are far below the diffraction limit of the probe we used ($\lambda/2NA = 1.21$ μm). The field of view is 6.8x6.8 μm$^2$.

Firstly, we simulated conventional point scan imaging using the Gaussian PSF presented in Fig. 2h. The full width at half maximum (FWHM) of the Gaussian PSF corresponds to the diffraction limit of the MM fiber. The spatial power spectrum is presented in Fig. 1c by the light-blue filled area. We calculated the overlap integral between the Gaussian PSF and the sample for different positions of the pump beam. A total of 4900 values were calculated for each sample. The resulting images are presented in Fig. 2b and 2e for 390 nm and 480 nm feature sizes, respectively. The conventional microscopy doesn't resolve any sub-diffraction details.

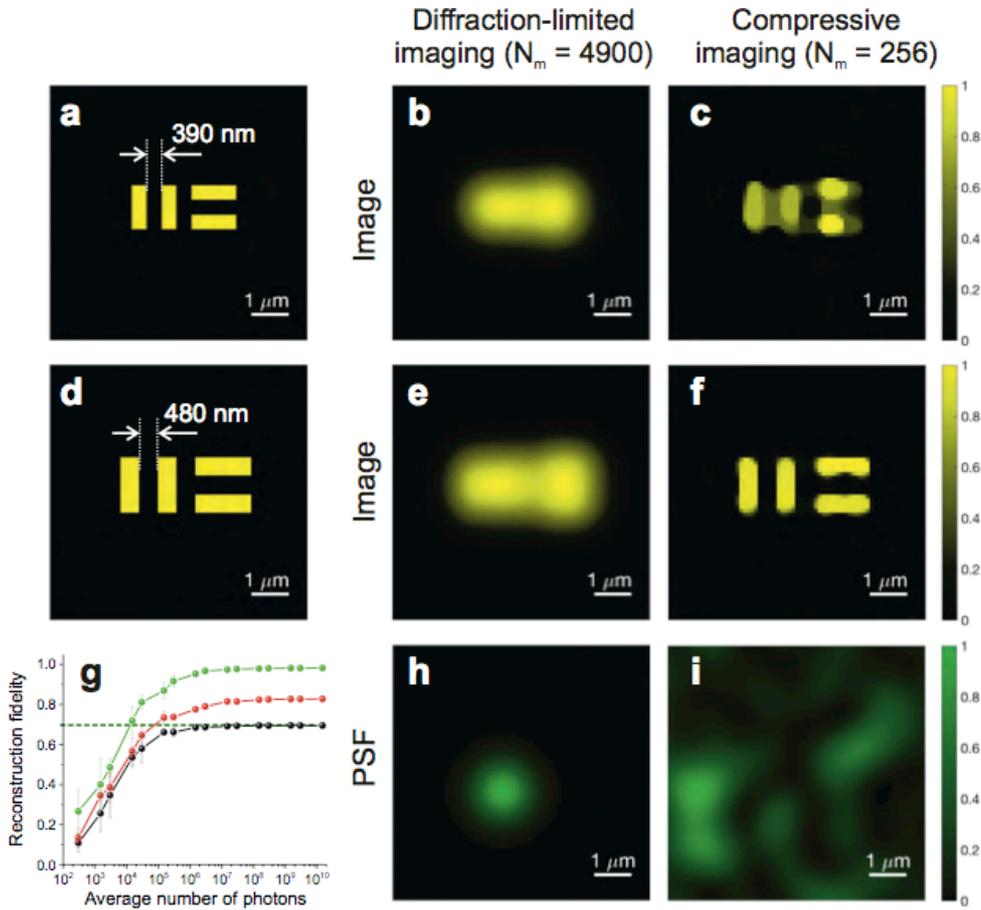

**Fig. 2 | Theoretical comparison of conventional imaging and super-resolution endo-microscopy. a, d,** Samples with 390 nm (**a**) and 480 nm (**d**) features. **b, e** Conventional diffraction-limited point scan imaging of the sample with 390 nm (**b**) and 490 nm (**e**) features. The total number of measurements $N_m$ = 4900. **c, f,** Super-resolution endo-microscopy of the sample with 390 nm (**c**) and 490 nm (**f**) features. The total number of measurements $N_m$ = 256. **g,** The fidelity of compressive endo-microscopy as a function of an average number of 'detected' photons in the presence of shot noise for 390 nm (black line), 480 nm (red line) and 580 nm (green line) features. The dashed green line represents a level of the fidelity, at which sub-diffraction details could be resolved. **h,** Gaussian PSF used for conventional point scan imaging modality. **i,** Example of an experimentally measured and low-pass filtered speckle-based PSF that was used for super-resolution compressive endo-microscopy. PSFs presented in **h** and **i** have the same cutoff frequency of $2NA/\lambda$.

Secondly, we simulated super-resolution compressive imaging through a MM fiber. As a sensing basis, we used the experimentally measured and low-pass filtered speckle patterns. An example is presented in Fig. 2i. We calculated the response for each speckle pattern of our sensing basis. A total of only 256 values were calculated for each sample. The collected data was scaled in such a way to represent the number of 'detected' photons. We added shot noise by generating random numbers from the Poisson distribution specified by the number of photons. As a first step, we did analyze a relatively low noise level: the average number of photons was $10^{10}$. To retrieve a 4900 pixels image from only 256 measurements we used a reconstruction procedure described in the Methods section. The acquired images are presented in Fig. 2c and 2f for 390 nm and 480 nm feature sizes, respectively. In contrast to conventional imaging, all sub-diffraction details are resolved even if the distance between them is more than 3 times smaller than the diffraction limit. Spatial power spectra of the acquired images resemble those of the original sample and spread way beyond the cutoff frequency of our optical system ($\nu_{cutoff}$), see Supplementary Fig. S3.

**Robustness to noise**

We analyzed the quality of super-resolution compressive endo-microscopy for different noise levels. We used three samples with sub-diffraction features separated by 390 nm (see Fig. 2a), 480 nm (see Fig. 2d) and 580 nm. We simulated super-resolution compressive endo-microscopy by using different average numbers of 'detected' photons per measurement, ranging from $10^2$ to $10^{10}$ (leading to a different signal-to-noise ratio according to Poisson statistics). Supplementary Movie 1 demonstrates how the acquired image improves with an increase in the number of detected photons for the 480 nm feature size. For hundreds of photons the image contains only noise, however at approximately $10^5$ photons all four stripes start to be resolved. Finally, from about $10^8$ photons we get a good image presented in Fig. 2f.

For each acquired image, we retrieved the reconstruction fidelity by calculating the cross-correlation coefficient, $k_s$, between the recorded image and the original sample (see Methods). The fidelities as a function of the number of photons for 390 nm, 480 nm and 580 nm features are presented in Fig. 2g by black, red and green color, respectively. The dashed line in Fig. 2g represents a level at which we can observe sub-diffraction details. The graph shows a similar behavior for different samples with the main distinction in the converging quality: the 580 nm features can be imaged with a high fidelity, whereas the 390 nm features can just be resolved.

**Simulation with low-pass filtered speckle patterns: beyond the Nyquist limit**

To analyze the super-speed property of our super-resolution endo-microscopy, we address the following questions: how many measurements are actually required to resolve sub-diffraction features of our sample and how are the imaging speed and the resolution related to each other. In our simulations, we kept the general structure of interest the same (see Fig. 2d) and increased the FOV and the sparsity by adding zero pixels. The total length of each dimension of the FOV varied from 6.4 μm to 10.6 μm and the total number of pixels $N_p$ from 4356 to 12100. For each sample size, we simulated the super-resolution approach to endo-microscopy for a different total number of experimentally measured speckle patterns, $N_m$ from 50 to 256. We did simulations for the relatively low noise case, in which the average number of 'detected' photons is $10^{10}$. It is important to mention that we always kept the number of measurements far below the Nyquist limit with a minimum of 0.4% to a maximum of 6% of the information and the sample contained features 2.5 times smaller than the diffraction limit of the fiber.

For each acquired image, we characterized the reconstruction fidelity by the cross-correlation coefficient, $k_s$, between the acquired image and the original sample. Figure 3a shows the image quality as a function of the total number of pixels $N_p$ and the total number of measurements $N_m$. Low values of the cross-correlation coefficient (blue color) represent images with a diffraction-limited resolution. An example is presented in Fig. 3b ($N_m = 100$ and $N_p = 6400$). A cross-correlation coefficient of more than approximately 0.7 (light blue color) corresponds to an image that still resolves sub-diffraction features of our object. An example is presented in Fig. 3c ($N_m = 160$ and $N_p = 6400$). Yellow color ($k_s$ close to 1) represents a very good reconstruction: An example for $N_m = 250$ and $N_p = 6400$ is presented in Fig. 3d. The total number of measurements $N_m$ that is needed for sub-diffraction imaging is linearly proportional to the total number of pixels $N_p$ (white dashed line), while $N_m \ll N_p$. The cross-sections over two stripes in the images in subfigures b, c and d are presented in Fig. 3e by the blue, red and green line, respectively. The cross-section of the image for standard diffraction-limited microscopy is presented in Fig. 3e by the filled light-blue area. The recorded images of the sample with 480 nm feature size as a function of the number of measurements for a total number of pixels $N_p = 8100$ are presented in Supplementary Movie 2. We can see the improvement of the image quality with an increase of the number of measurements.

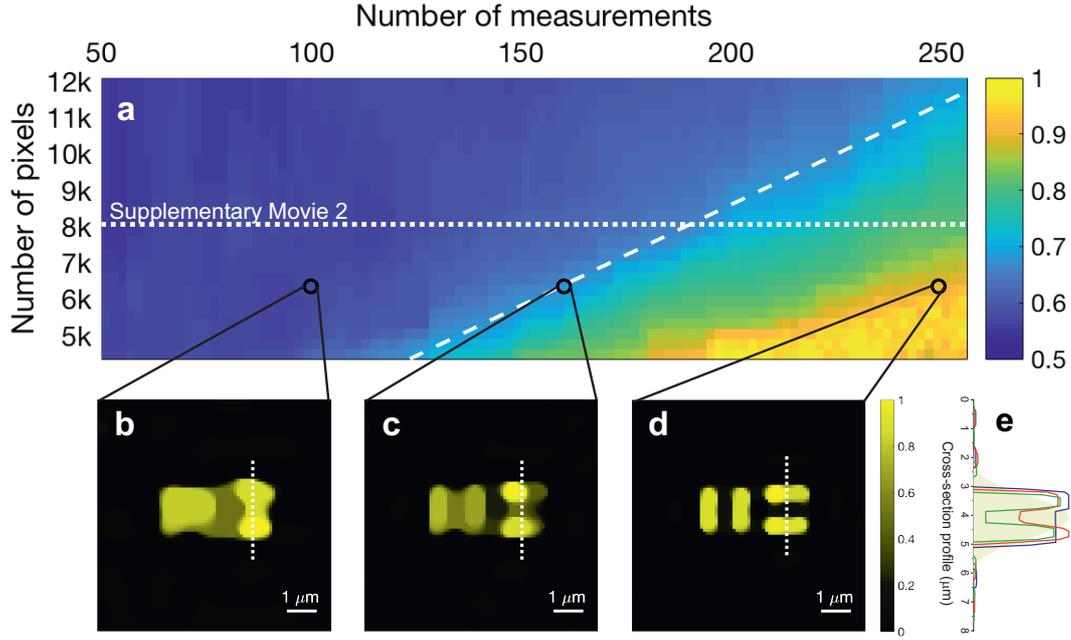

**Fig. 3 | Theoretical analysis of the trade-off between imaging speed and resolution below the diffraction limit**. **a,** Fidelity of the super-resolution compressive endo-microscopy as a function of the total number of pixels in the FOV ($N_p$) and the total number of measurements ($N_m$) for a sample with a feature size of 480 nm. The white dashed line shows that $N_m$ needed for sub-diffraction imaging is linearly proportional to $N_p$. The recorded images of the sample for $N_p = 8100$ (white dotted line) are presented in Supplementary Movie 2 as a function of the number of measurements. **b-d,** Results of the compressive endo-microscopy simulated by using experimentally measured and low-pass filtered speckle patterns on the MM fiber output and in the presence of shot noise for the sample with $N_p = 6400$: **b,** $N_m = 100$, **c,** $N_m = 160$; **d,** $N_m = 250$. **e,** Cross-sectional plots along the dashed lines in (b), (c) and (d) are presented by blue, red and green lines, respectively. The filled yellow area represents a cross-section of a conventional diffraction-limited image.

**Experimental demonstration of endo-microscopy beyond the Abbe and Nyquist limits**

In the first set of measurements, we demonstrated super-resolution and super-speed compressive endo-microscopy through the first MM fiber with NA = 0.22. The details of the imaging procedure are described in the Methods section. Similar to our simulations, we performed only 256 measurements and reconstructed the full FOV consisting of more than 4000 pixels. The result is presented in Fig. 4a and a zoomed-in area of interest is presented in Fig. 4b. We compared the results of the proposed approach with the state-of-the-art method of MM fiber imaging based on complex wavefront shaping[12], where the image is retrieved after point by point scanning of a focused laser beam along the MM fiber output facet. For wavefront shaping based endo-microscopy, we used the same setup and the same fiber probe to maintain the same numerical aperture and the resolution of the system. We did not use the same sample, because the small spheres demonstrated pronounced photobleaching of the fluorescent signal that may influence the performance of the second measurement. Instead we used a similar sample consisting of spheres with exactly the same diameter (standard deviation was 9 nm). The result (a zoom-in of the area of interest) is presented in Fig. 4c.

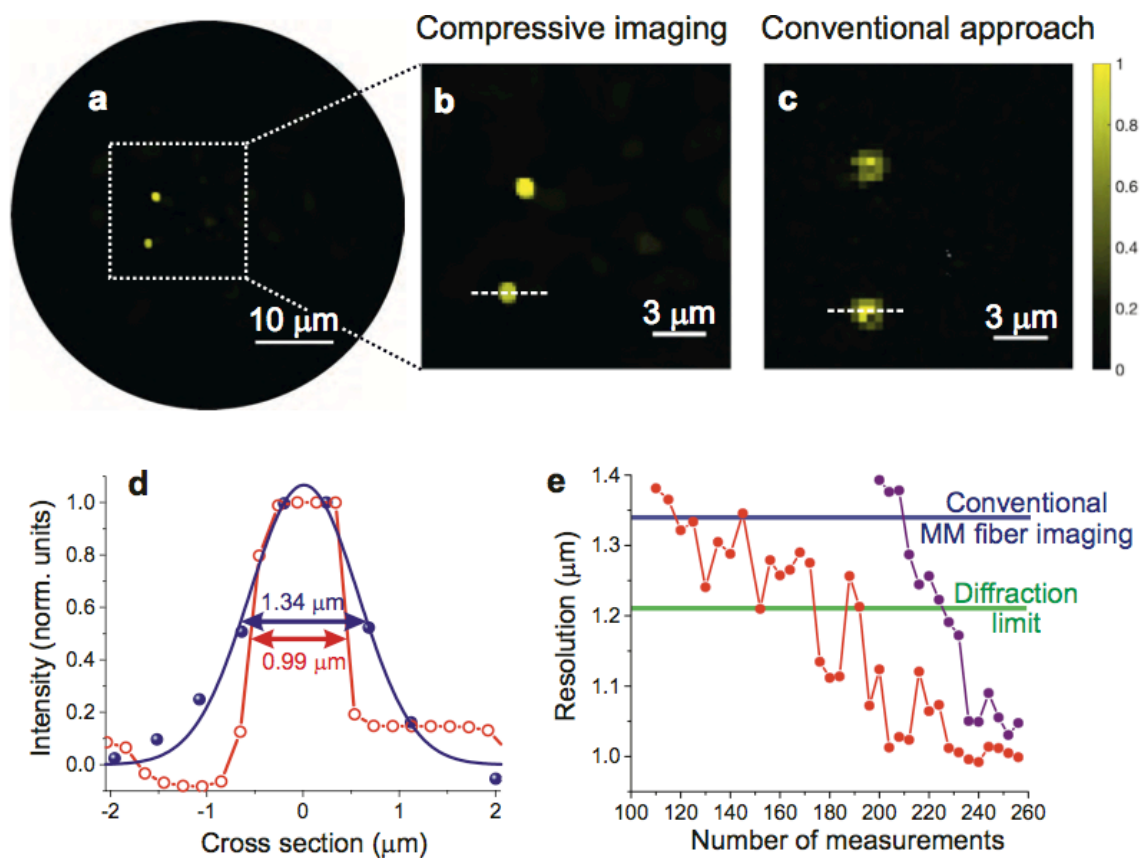

**Fig. 4. | Super-resolution and super-speed compressive endo-microscopy through the MM fiber with NA = 0.22. a-c,** Images of two 450-nm-diameter fluorescent spheres through the MM fiber probe by **a, b,** proposed super-resolution endo-microscopy (full FOV is presented in (a) and zoomed-in area of interest is presented in (b)) and **c,** state-of-the-art raster scan wavefront shaping based endo-microscopy. **d,** Cross-sectional plots along the dashed lines in b (open red circles) and c (filled blue circles). The blue line is a Gaussian fit to the data, the red line interpolated the data. **e,** Resolution of compressive endo-microscopy as a function of the total number of measurements for two fluorescent spheres is presented by different colors. The diffraction limit is presented by the solid green line and the resolution achieved in the same setup by the state-of-the-art method of endo-microscopy is presented by the solid blue line.

A cross-section of the fluorescent sphere imaged with the conventional laser scanning endo-microscopy approach is presented in Fig. 4d by blue circles. It follows a Gaussian shape (blue solid line) with a FWHM = 1.34 μm, which is slightly above the diffraction limit due to inaccuracies in the wavefront shaping and the distance between the fiber facet and the sample. The cross-section of the fluorescent sphere imaged with the compressive super-resolution endo-microscopy is presented in Fig. 4d by open red circles. In contrast to the state-of-the-art method, it has very sharp edges and doesn't follow the conventional Gaussian shape. The FWHM of 0.99 μm was estimated using interpolation. The measured sphere diameter is 20% below the diffraction limit of our optical system. Simultaneously, we used only about 5% of the information, meaning the imaging speed was 20 times faster than would be needed for scanning based imaging with any other technique.

In the second set of measurements, we experimentally analyzed the intrinsic trade-off between the spatial resolution and the imaging speed. We repeated the compressive endo-microscopy experiment for a different number of measurements varying from 110 to 256. The results are present in Fig. 4e where the experimentally measured resolution of the proposed approach for two fluorescent spheres is plotted as a function of the number of measurements. The diffraction limit is presented by the solid green line and the resolution achieved in the same setup by the state-of-the-art method of endo-microscopy is presented by the solid blue line. We see that by reducing the number of measurements and keeping the same sample we reduce the resolution, as was predicted by our simulations.

Equivalent measurements were taken for an MM fiber with a core of 105 um and NA = 0.1. This fiber has a diffraction limit of 2.66 μm and supports approximately the same total number of modes as the first one. Consequently, despite the twice bigger FOV we expect a similar resolution improvement according to our simulations. To demonstrate this we compare the performance of the proposed approach (Fig. 5a) with state-of-the-art endo-microscopy (Fig. 5b). We used the same setup and the same sample for both measurements.

Cross-sectional plots along the dashed lines in Fig.5a and 5b are presented in Fig. 5c by blue triangles and red circles, respectively. The FWHM for the proposed endo-microscopy approach is 1.93 μm, which is about 25% below the diffraction limit of our optical system. Simultaneously, we performed only 225 measurements, meaning the imaging speed was 20 times faster than would be needed for scanning based imaging. The trade-off between the spatial resolution and the imaging speed is presented in Fig. 5d, where the experimentally measured resolution averaged over four fluorescent spheres is plotted as a function of the total number of measurements. The error bars represent the standard deviation. The diffraction limit is presented by the solid green line.

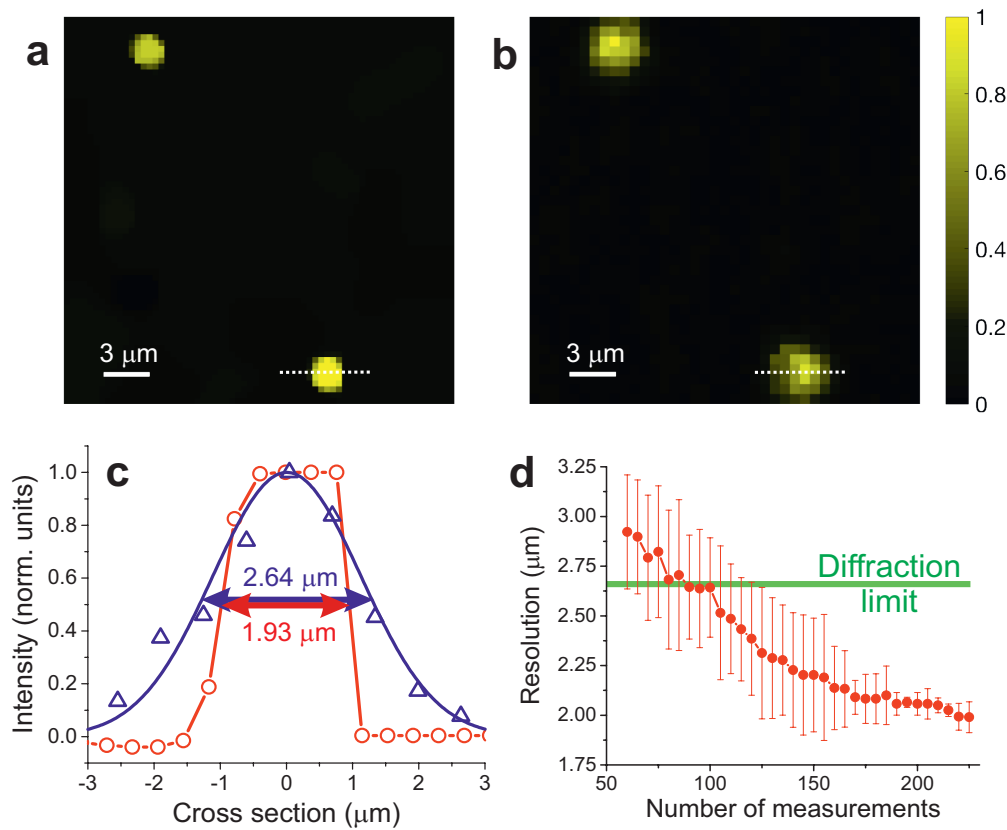

**Fig. 5. | Super-resolution and super-speed compressive endo-microscopy through the MM fiber with NA = 0.1. a-b,** Images of two fluorescent spheres through the MM fiber probe by **a,** proposed super-resolution compressive endo-microscopy and **b,** state-of-the-art raster scan wavefront shaping based endo-microscopy. **c,** Cross-sectional plots along the dashed lines in a (open red circles) and b (open blue triangles). The blue line is a Gaussian fit to the data, the red line interpolated the data. **d,** Resolution of compressive endo-microscopy as a function of the total number of measurements averaged over 4 fluorescent spheres. The error bars represent the standard deviation. The diffraction limit is presented by the solid green line.

**Improving the resolution by reducing the FOV and the noise level**

Our theoretical analysis shows that the resolution depends on the total number of measurements and could be significantly improved by reducing the total FOV for the same number speckle patterns forming the basis. In the final set of measurements, we experimentally demonstrated this effect by restricting the image reconstruction to a smaller area of the fiber. We used a fiber with NA = 0.1 (diffraction limit of 2.66 μm) and core diameter of 105 μm, but we reconstructed the image only in a 30x30 micron area on the fiber facet, while we made sure no fluorescent spheres were present outside the reconstructed area. In this manner we artificially mimicked a fiber with a small FOV. We used bigger fluorescent spheres (approximately 1.5 μm in diameter) to reduce the noise level.

The bright field image of the sample is presented in Fig. 6a. The separation between the left and the middle spheres is about 1 μm and the separation between the middle and the right spheres is less than 3 μm. We repeated the experimental procedures described in the previous section. The image of the sample obtained with state-of-the-art endo-microscopy based on wavefront shaping is presented in Fig. 6b (zoomed-in area of interest). We see that the left and the middle spheres are completely unresolvable, the middle and the right spheres are hardly resolved, as expected. We used the same fiber probe and the same sample for super-resolution endo-microscopy experiments. The result (zoomed-in area of interest) is presented in Fig. 6c. All 3 spheres are fully resolved.

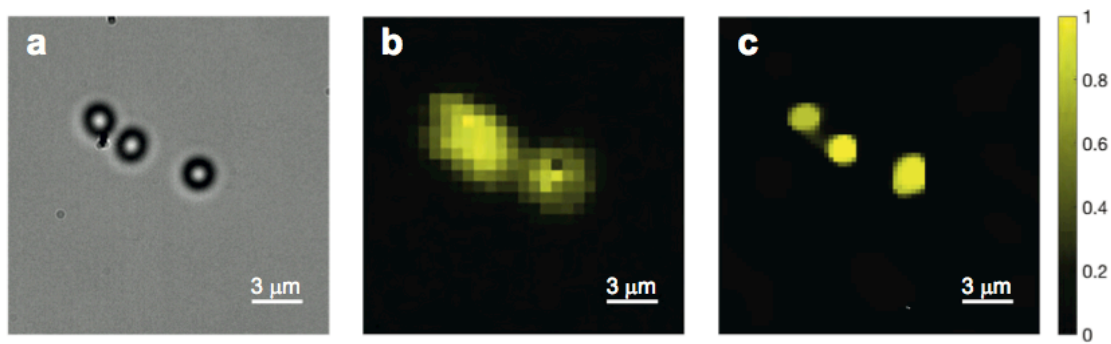

**Fig. 6. | Resolution improvement of compressive endo-microscopy by reducing the FOV and the noise level. a,** Bright-field image of the fluorescent spheres. The separations between the left and the middle spheres and between the middle and the right spheres are about 1 μm and less than 3 μm, respectively. The diffraction limit of the system is 2.66 μm. **b-c,** Images obtained through a MM fiber probe in endoscopic regime by state-of-the-art raster scan wavefront shaping based endo-microscopy (**b**), and proposed super-resolution compressive endo-microscopy with reduced FOV (**c**).

**Discussion**

We have demonstrated that the compressive sensing approach in combination with a MM fiber probe can provide imaging beyond the Abbe and Nyquist limits simultaneously in an endoscopic configuration. Our simulations show that the spatial resolution can be improved more than 3 times below the diffraction limit and simultaneously the speed more than 20 times faster than dictated by the Nyquist limit. Supplementary Movie 3 represents an experimentally acquired images as a function of the total number of measurements for super-resolution compressive endo-microscopy (left) and for state-of-the-art point scan endo-microscopy based on wavefront shaping (right). The experiments were done on the same setup, same level of signal and the same fiber probe. This movie clearly demonstrates the advantages of super-resolution compressive endo-microscopy: a significantly lower number of measurements is needed to get a full FOV image with resolution below the diffraction limit. Whereas in the case of the conventional approach, the same number of measurements allows to visualize only a small fraction of the FOV with a diffraction-limited resolution at best.

We addressed practical limitations of the resolution improvements, such as noise, stability and reproducibility of basis vectors, orthogonality of the sampling basis and imaging speed (number of measurements). We concluded (see Fig. 3, Fig. 4e and Fig. 5d) that the imaging speed and the resolution are closely related: the better resolution we want to get the more measurements we need to perform. We carefully analyzed an effect of shot noise (see Fig. 2g and Supplementary Movie 1). The estimation of the fundamental spatial resolution limit as a function of all the parameters is a subject of further investigation. However, it is important to note that we demonstrate that the spatial resolution of the proposed compressive endo-microscopy is not limited to a 2 times improvement.

Remarkably, the new super-resolution approach of endo-microscopy does not use any specific properties of the fluorescent label typical for many super-resolution techniques such as depletion (STED) or stochastic activation of the molecular fluorescent (PALM). The main limiting factor related to the sample is the number of registered photon: the sample should be bright enough to provide more than $10^4$ multiplied by $N_m$ photons. The method in present form is not suitable for single molecule imaging due to the fact that conventional dye molecules usually only emit $10^6$ photons before being bleached. The super-resolution endo-microscopy is ready to be implemented in a variety of fluorescent and label-free modalities of microscopy.

**Materials and methods**

**Experimental setup.** All the experiments were performed using a standard step-index MM fibers [Thorlabs, FG050UGA and FG105LVA] with a silica core diameter of 50 μm, NA = 0.22 and 105 μm, NA = 0.1. The fibers were not straight, a slight curve was introduced to improve cross-mode coupling, the length of each fiber was approximately 20 cm. The detailed sketch of the experimental setup is presented in Supplementary Fig. S1. As a pump, we used the continuous-wave linearly polarized second harmonic output of a Nd:YAG laser [Cobolt Samba] with a wavelength of 532 nm and power below 1 mW. As a laser scanning element for the endo-microscopy, we used a spatial light modulator:

1920x1200 Vialux V4100 digital micromirror device (DMD). This choice stems from the ability of the DMD to perform phase control that is required for the comparison of our new approach with the state-of-the-art wavefront shaping based method of high-resolution endo-microscopy. The beam was expanded by the telescope to match the area of the DMD. Two lenses were placed in a 4f configuration to image the DMD on the back focal plane of a 20x (NA = 0.4) or 10x (NA = 0.25) objective that coupled the light into the MM fiber. A pinhole in the Fourier plane blocked all but the 1$^{st}$ diffraction order, encoding the desired phase or amplitude distribution. The DMD was used to control the spatial phase (for state-of-the-art endo-microscopy) or the spatial amplitude (for super-resolution compressive endo-microscopy) of a beam on the input facet of the fiber. A 50x (NA = 0.75) or 40x (NA = 0.6) objective was used to collect light from the fiber output (transmission side). The camera imaged the output facet of the MM and was used for pre-calibration. The same MM fiber was used to collect the total fluorescent response and propagate it in the backward direction to the detection system. We used a dichroic mirror to separate the fluorescent signal and an additional Notch filter to remove pump light. For detection, a standard avalanche photodiode [Thorlabs, APD410A2] in combination with a low-noise preamplifier [Stanford Research] was used.

As a sample, we used fluorescent microparticles (PS-FluoRed-Fi144-2) with a diameter of 450 ± 9 nm and fluorescent microparticles (Nile red) with a diameter 1.1 – 1.5 μm. Microparticles were randomly distributed over a glass substrate.

**State-of-the-art wavefront shaping based endo-microscopy.** During the pre-calibration step, the complex wavefront shaping algorithm[13] was used to create tightly focused spots on the fiber output. The phase masks corresponding to focal spots at different positions on the output fiber facet were calculated and stored. After the wavefront shaping procedure, the sample was placed against the fiber facet as close as possible without touching. The sample image was acquired in the endoscopy configuration by sequentially applying the recorded phase masks to the DMD and detecting the total fluorescent signal. As a result, the pixel-by-pixel image of the sample was reconstructed and about 4000 measurements are required to provide an image with diffraction-limited resolution over the full FOV, according to the Nyquist-Shannon sampling theorem.

**Compressive endo-microscopy.** The pre-calibration procedure consisted of recording a set of 256 speckle patterns on the output facet of the MM fiber for different positions of the focal spot on the fiber input facet organized in a 16x16 square lattice. The background signal that corresponded to each speckle pattern was also recorded. After the pre-calibration, the sample was placed against the output fiber facet as close as possible without touching. Super-resolution and super-speed compressive endo-microscopy was performed by scanning the focal spot at the input fiber facet in the same way as during the pre-calibration procedure. The total fluorescent response for each speckle pattern collected by the MM fiber probe was measured by a single-pixel detector. A total of 60 to 256 measurements were taken in different experiments. To retrieve an image, we used the reconstruction procedure described the following subsection.

**Image reconstruction.** The fundamental problem is to recover a vector $x \in \mathbb{R}^n$ from data $y$: $y = Ax + z$, where A is a known m×n 'sensing matrix', which is used in the measurements, $z$ is unknown error term, $n$ is the number of pixels, $m$ is the number of 'measurements'. We are interested in the case where $m \ll n$, and the distance between two neighboring pixels $\Delta n$ is smaller than the Abbe diffraction limit $\lambda/2NA$, where $\lambda$ is the wavelength and NA is a numerical aperture of our optical system.

We used an established procedure as the solution to the $l_1$ minimization problem[44]. Among all objects consistent with the data, we pick the one whose coefficient sequence has a minimal $l_1$ norm:

$$\min_{\tilde{x} \in \mathbb{R}^n} \|\tilde{x}\|_{l_1} \ subject\ to\ A\tilde{x} = y \tag{1}$$

It is a standard linear programming problem that can be solved in polynomial time using one of the many existing software packages. We used an open software algorithm '$l_1$ magic' from Stanford.edu[45].

**Cross-correlation coefficient.** We calculate the correlation coefficient $r$ between any two measured speckle patterns, by using the following equation:

$$r = \frac{\sum (a - \bar{a})(b - \bar{b})}{\sqrt{\sum (a - \bar{a})^2 \sum (b - \bar{b})^2}} \tag{2}$$

where $a$ and $b$ are images of speckles on the output facet of a MM fiber, $\bar{a}$ and $\bar{b}$ are their means. A total of 256 speckle patterns were created by scanning a focused input beam over 256 points organized in a square lattice (16x16) on the input fiber facet. Correlation coefficients are calculated for all pairs of speckle patterns and the results are presented in Fig. 1b.

We also use cross-correlation coefficient to characterize the fidelity of the acquired image. The reconstruction fidelity is calculated as the cross-correlation coefficient between the recorded image and the original sample. In that case, $a$ and $b$ are the recorded image and the original sample, respectively, $\bar{a}$ and $\bar{b}$ are their means.

**Acknowledgements**
We acknowledge the financial support of the Nederlandse Organisatie voor Wetenschappelijk Onderzoek (NWO) (VENI 15872).

**Conflict of interest** The authors declare that they have no conflict of interest.
The data that support the findings of this study are available from the corresponding author upon reasonable request.

# Supplementary material

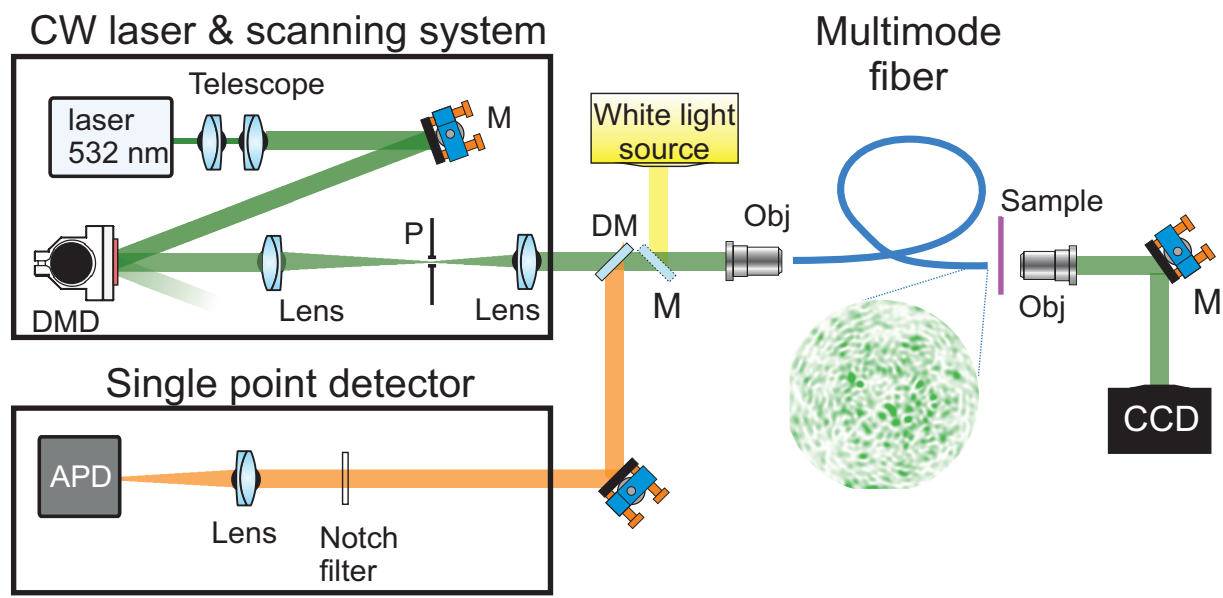

**Fig. S1 | Experimental setup.** The experimental setup consists of 3 main components: a continuous-wave (cw) laser source with a scanning system, a MM fiber probe, and a single-point detector. Pump light is scanned across the fiber input facet creating different illumination patterns on the fiber output. Proposed super-resolution and super-speed compressive endo-microscopy is performed by using the scanning system only for amplitude control. State-of-the-art wavefront shaping based point scan endo-microscopy is performed by using the scanning system for spatial phase control only. The total fluorescent response from the sample is collected by the same fiber probe, propagated back and measured by the single point detector. Bright-field microscopy is used for reference. CCD camera is used for pre-calibration. DMD, digital micromirror device; M, mirrors; P, pinhole; DM, dichroic mirror; Obj, objectives; APD, avalanche photodiode.

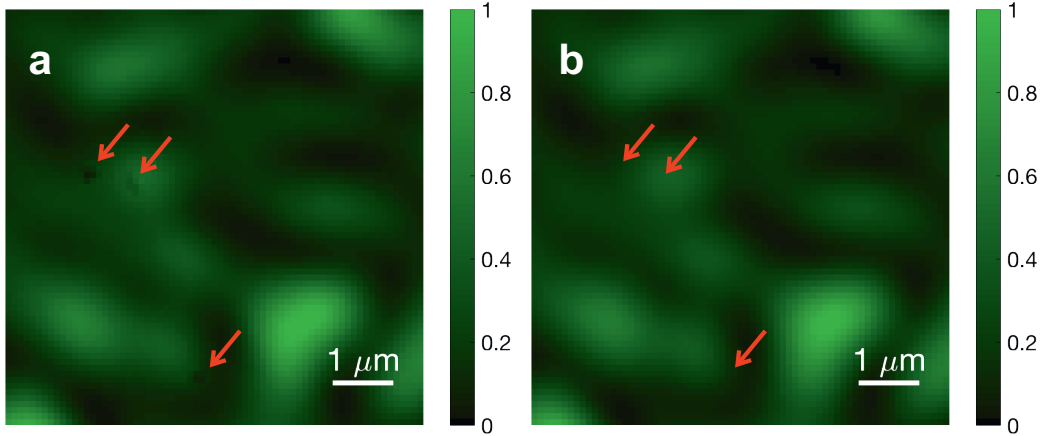

**Fig. S2 | Analysis of the measured speckle patterns.** Zoomed-in camera image of the speckle pattern generated by the MM fiber **a,** before and **b,** after low-pass filtering by cutting the spatial power spectrum at $\nu_{cutoff} = 2NA/\lambda$. Red arrows indicate the presence of dead pixels on the camera sensor that lead to high-frequency noise. After low-pass filtering, there are no high-frequency components in speckle PSF.

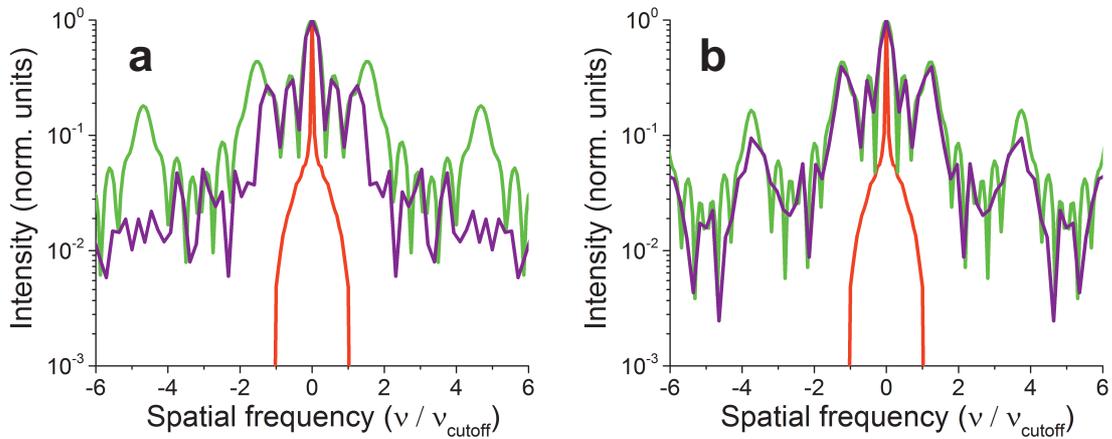

**Fig. S3 | Spatial power spectra cross-sections. a,** Cross-section of spatial power spectra of the sample with feature size of 390 nm. **b,** Cross-section of spatial power spectra of the sample with feature size of 480 nm. Red lines represent the average low-pass filtered spatial power spectra of speckle patterns generated in the MM fiber and used for imaging. Green lines represent spatial power spectra of the original samples. Violet lines show spatial power spectra of the image acquired by super-resolution compressive endo-microscopy. The cutoff frequency is equal to $2NA/\lambda$.

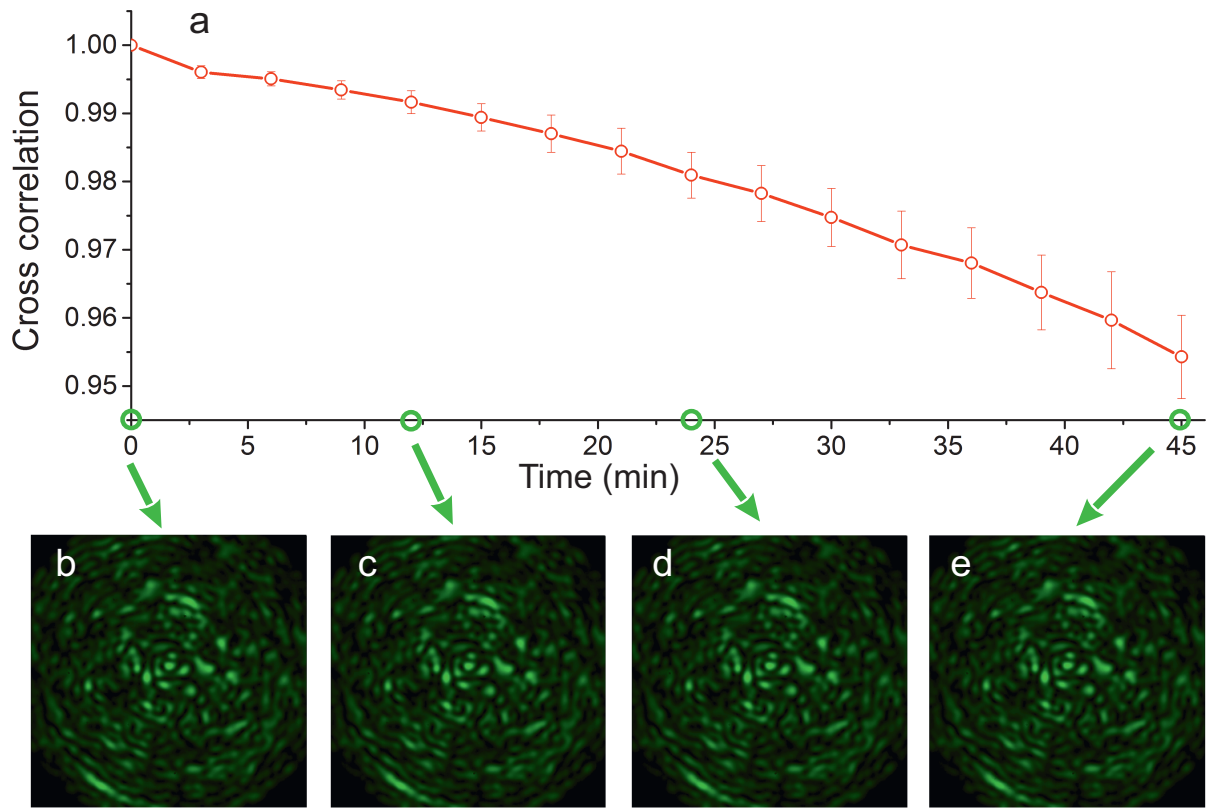

**Fig. S4 | Reproducibility and stability of speckle patterns. a**, Cross-correlation coefficient between the speckle pattern measured at zero time and the speckle pattern measured after sequentially switching to 99 other positions, averaged over 100 positions of the focal spot on the fiber input facet. **b-e** Examples of the recorded speckle patterns for a particular position of the focal spot on the fiber input facet for 0 (b), 12 (c), 24 (d) and 45 (e) minutes.